\documentclass[amsfonts,amssymb,amsmath,showpacs,byrevtex]{revtex4}
\usepackage{amsthm}
\newtheorem{cor}{Corollary}
\newtheorem*{thm}{Theorem}

\begin{document}

\title{Finite Size Universe or Perfect Squash Problem}
\author{Ludwik Turko}
\email{turko@ift.uni.wroc.pl}

\affiliation{Institute of Theoretical Physics, University of Wroc{\l}aw,\\ Pl. Maksa
Borna 9, 50-204  Wroc{\l}aw, Poland}

\date{April 21, 2004}

\begin{abstract}
We give a physical notion to all self-adjoint extensions of the operator $id/dx$ in the
finite interval. It appears that these extensions realize different non-unitary
equivalent representations of CCR and are related to the momentum operator viewed from
different inertial systems. This leads to the generalization of Galilei equivalence
principle and gives a new insight into quantum correspondence rule. It is possible to get
transformation laws of wave function under Galilei transformation for any scalar
potential. This generalizes mass superselection rule. There is also given a new and
general interpretation of a momentum representation of wave function. It appears that
consistent treatment of this problem leads to the time-dependent interactions and to the
abrupt switching-off of the interaction.
\end{abstract}

\pacs{03.65.-w, 03.65.Ca, 03.65.Db, 03.65.Ge, 02.30.Tb} %
%%%%%%%%%%%%%%%%%%%%%%%%%%%%%%%%%%%%%%%%%%%%%%%%%%%%%%%%%%%%%%%%%%%%%%%%%%%%%%%%%%%
%03.65.-w Quantum mechanics %
%03.65.Ca Formalism %
%03.65.Db Functional analytical methods %
%03.65.Ge Solutions of wave equations: bound states %
%02.30.Tb Operator theory %
%%%%%%%%%%%%%%%%%%%%%%%%%%%%%%%%%%%%%%%%%%%%%%%%%%%%%%%%%%%%%%%%%%%%%%%%%%%%%%%%%%%
\maketitle
\section{\label{sect:intr}Introduction}
A square well potential, although this is the simplest analytically solvable quantum
model, can be used as a tool to investigate more involved quantum peculiarities. It was
used recently for such different phenomena as quantum fractals \cite{berr,wojc}, quantum
chaos \cite{Hu:1999bj} or wave-function revivals \cite{aron,robin1999}. It can also be
used as as approximation to experimentally realized semiconductor quantum well lasers or
micromaser cavities with atomic rubidium \cite{verd,arakyari}.

Schr\"odinger equation with a square well potential can also be considered as a model of
a quantum squash. An infinite well corresponds to perfectly rigid and perfectly resistant
side walls. A finite square well potential corresponds to perfectly rigid but not
perfectly resistant side walls --- a high energy squash ball breaks through the wall. A
physicist is here like a passive player. He or she (=(s)he) can use a racket only as a
measurement apparatus --- to register the energy or the momentum of the ball.

A simplicity of the model may be misleading. A closer inspection (see \emph{e.g.}
\cite{antoin2001gm,bonneau:2001zq,garb}) shows  that the infinite potential well has
mathematical traps which, when neglected, lead to contradictions or misinterpreted
results.

The aim of this paper is to study physical consequences of different self-adjoint
extensions of the ``momentum" operator for a quantum squash. The ``momentum" means here
the differential operator $-i\hbar\nabla$. In the case of square integrable functions on
$\mathbb{R}^n,\ (n=1,2,3)$ this operator is self-adjoint --- so it is interpreted as the
momentum operator. A situation is much more involved for a particle in a box. There are
infinitely many self-adjoint extensions of the ``momentum" operator. Different extensions
correspond to different boundary conditions of functions from the domain of the operator
and they have different spectra. A question arises which of these extensions is the
physical momentum operator and what are physical notions of other self-adjoint extensions
of the operator $-i\hbar\nabla$?

It will be shown that all those self-adjoint extensions have physical meanings. They are
closely related to the Galilei transformed reference frames moving with different
velocities with respect to the primary frame. The primary frame is chosen as the frame
with a time-independent potential. This means that our squash play does not move on the
squash field. When (s)he changes this passive strategy and starts to run (with a constant
velocity of course) then (s)he observes shifted momenta of the squash ball. This picture
is in a perfect agreement with a physical ``classical" intuition. It appears also that
when (s)he solves corresponding Schr\"odinger equation then a transformed wave function
behaves according to projective representations of the Galilei group.

A r\^ole of projective representations of the Galilei group is well established since
long \cite{Inonu:1952ct,bargm,hamerm,levy:1963,levy:1974,giul:1996}. All results were
then obtained with Galilei-invariant potentials. For one particle this was equivalent to
a free-particle case. It appears that basic results, the Bargmann superselection rule
including, can be reproduced for any potential.

Finally, we are going to clarify the momentum representation puzzle. Let us consider a
squash player confined to the finite region bounded by perfectly rigid and perfectly
resistant side walls. For the player this squash-room is like a finite Universe. The
spectrum of the momentum operator is discrete in this Universe. Since according to basic
rules od quantum mechanics the only possible results of the momentum measurement are
eigenvalues of the corresponding observable the momentum distribution should be discrete
one. However, there is a common procedure (see \emph{e.g.} \cite{land,coh}) to take the
Fourier integral transform of the wave function. This Fourier transform is interpreted as
the momentum representation. This inconsistency was also observed in \cite{antoin2001gm}
but authors didn't push the problem further.

One can show that both momentum representations have well established physical
interpretations, although both describe different physical situations. The Fourier
integral transformation of the wave function is simply related to the abrupt switch-off
of the potential. As the infinite square well potential can be used as a model for the
perfect squash so the Fourier integral of the harmonic oscillator wave function can be
used for the quantum sling theory.

We begin by consideration of the notion of momentum distributions. It appears that a
momentum distribution of the wave function understood as a Fourier transform is directly
related to the solution of Schr\"odinger equation with a time-dependent interaction. The
Fourier transform $\Phi(\vec p,t)$ of the wave function $\Psi(\vec r,t)$ is the
probability amplitude to measure at time $t>0$ the momentum $\vec p$ when an interaction
was switch-off at $t=0$. This gives also a new insight into David-Goliath fight, as is
presented in Section \ref{subs:david}.

In Section \ref{sect:math} some  necessary mathematical preliminaries are given. These
are related to self-adjoint extensions of differential operators $id/dx$ and $d^2/dx^2$.
This material does not pretend to give a new insight into the problem, but collects some
mathematical facts not always known to the physical community. An analysis of stationary
solutions of the infinite square well potential is given as an example in Section
\ref{subs:sgm:repr}.

Section \ref{sect:mom:mv} deals with a physical interpretation of self-adjoint extensions
of the operator $-id/dx$ in the Hilbert space of square integrable functions on a finite
interval. First an analysis of the notion of the quantum momentum observable is
performed. An operator can be identified with the physical momentum only if it transforms
under Galilei transformation similarly to the classical momentum. This assumption allows
to add physics to all self-adjoint extensions of the operator $-id/dx$. These extensions
correspond to momenta seen by moving observers from different inertial systems. It will
be shown that those different extensions realize different non-unitary equivalent
representations of Canonical Commutation Relations.

A natural problem which arises at that moment is to find how different moving observers
see quantum mechanics from their systems. It is well known since papers of Bargmann,
In\"onu and Wigner \cite{Inonu:1952ct,bargm} that free Schr\"odinger equation is Galilei
invariant provided that wave function transforms under a projective representation of the
Galilei group. Section \ref{sect:schr:mv} deals with this problem and generalizes a
concept of Galilean covariance to any scalar potential. Now, a problem of the momentum
distribution is reexamined. Solutions of the infinite potential well are taken as
examples. It appears that momentum distributions are more tricky as it seemed before. In
particular, a mathematical identity
\[\sin x = \frac{1}{2i}(e^{ix}-e^{-ix})\]
is not so obvious in a quantum word. This is explained in Section
\ref{subs:mom:mv:well}.

Final conclusions are given in Section \ref{sect:concl}.

\section{\label{sect:momdstr}Momentum distributions}
It is a common knowledge that there is the discrete energy spectrum of a quantum particle
placed in an infinite square well potential. A one-dimensional potential of the form
\begin{equation}\label{well}
    U(x)=\begin{cases}
0\,, & \mbox{  for $0\leq x\leq a$}\,,\\
\infty\,, & \mbox{  for $x$ everywhere else}\,,
\end{cases}
\end{equation}
with boundary conditions
\begin{equation}\label{bound}
    \psi(0)=\psi(a)=0\,,
\end{equation}
leads to the solutions
\begin{equation}\label{sol}
\psi_N(x)=\left\{\begin{array}{cc}
 \sqrt{\frac{2}{a}}\sin{\frac{N\pi}{a}x}, & \mbox{  for $0\leq x\leq a$}\,,\\
0, & \mbox{  for $x$ everywhere else}\,,
\end{array}\right.
\end{equation}
where $N$ is an arbitrary positive integer.

Corresponding energy levels are
\begin{equation}\label{sol:en}
    E_N=\frac{\pi^2\hbar^2}{2 m a^2}N^2\,.
\end{equation}

Let us consider the Fourier integral of the wave function \eqref{sol}
\begin{equation}\label{four}
    \tilde\psi_N(k)=\frac{1}{\sqrt{2\pi}}\int\limits_0^a dx\,\psi_N(x)\,e^{-i k x}\,.
\end{equation}
One gets
\begin{equation}\label{sol:four}
 \tilde\psi_N(k)=-\sqrt{\pi a}\,\frac{2 N}{a^2 k^2-N^2\pi^2}\,e^{-i a k/{2}}
 \begin{cases}
    i\sin{\frac{a k}{2}}\,, & \mbox{ for $N$ even}\,,\\
    \cos{\frac{a k}{2}}\,, & \mbox{  for $N$ odd}\,.
\end{cases}
\end{equation}
This mathematical expression is usually (see \emph{e.g.} \cite{land,coh} and a lot of
other textbooks) interpreted as the physical momentum (with $p=\hbar k$) representation
of the wave function. According to such an interpretation the probability distribution of
the measurement of the momentum yielding a result between $p$ and $p+dp$ is
\begin{equation}\label{prob1}
\mathcal{P}_N(p)=\frac{4\pi a\hbar^3 N^2}{(a^2 p^2-\hbar^2 N^2\pi^2)^2}
\begin{cases}
    \sin^2{\frac{ap}{2\hbar}}\,, & \mbox{ for $N$ even} \,,\\
    \cos^2{\frac{ap}{2\hbar}}\,, & \mbox{  for $N$ odd}\,.
\end{cases}
\end{equation}
This gives an average value of the momentum equal to zero, and an average value of the
squared momentum is
\begin{equation}\label{momsqav}
  \langle p^2\rangle_N= \int\limits_{-\infty}^{+\infty}dp\,p^2 \mathcal{P}_N(p)=\frac{N^2
  \pi^2\hbar^2}{a^2}\,.
\end{equation}
This is in agreement (in an average) with \eqref{sol:en}. This is,
however, not an answer for the question about the squash ball
momentum. Besides that, there is a question about the energy
conservation: how is it possible to get any value of the momentum
in the state with a given value of the energy \eqref{sol:en}?

For the player in h(is)er finite Universe $0\leq x\leq a$ the only
allowed values of a momentum are those which are eigenvalues of
the corresponding self-adjoint observable. Using the trivial
identity
\begin{equation}\label{spr}
    \psi_N(x)=\sqrt{\frac{2}{a}}\sin{\frac{N\pi}{a}x} =
\frac{1}{2i}\sqrt{\frac{2}{a}}\left(e^{\frac{i\pi N}{a}x}-e^{-\frac{i\pi
N}{a}x}\right)\,,
\end{equation}
one gets a simple conclusion that allowed values of momenta are
$\pm \frac{N\pi\hbar}{a}$. This is of course in a perfect
agreement (not only in an average) with \eqref{sol:en}.

Such kind od contradictions led recently to conclusion \cite{antoin2001gm} that the
Fourier integral ``is just a mathematically equivalent version of the same object,
\emph{not} the momentum representation wave function."

As we'll see later, the using of Eq. \eqref{spr} as a plane waves
superposition is an oversimplification of the problem. There is,
however, a surprisingly simple answer to a question about the
physical notion of Eq. \eqref{sol:four}.
\subsection{\label{subs:mdistr:gen}General momentum distribution}
Let us consider a time-dependent hamiltonian
\begin{equation}\label{pot:tm}
\hat H=\begin{cases}
 -\frac{\hbar^2}{2m}\Delta + U(\vec r), & \mbox{  for $t\leq 0$}\,,\\
 -\frac{\hbar^2}{2m}\Delta, & \mbox{  for $t>0$}\,.
\end{cases}
\end{equation}
Let $\psi$ be any solution of the Schr\"odinger equation
\[i\hbar\frac{\partial\psi}{\partial t}=-\frac{\hbar^2}{2m}\Delta\psi + U(\vec r)\psi\,.\]
A general form of the free Schr\"odinger equation is a wave packet
\[\int d^3p\,g(\vec
p)\,e^{-i\frac{p^2}{2m\hbar}t}\,e^{\frac{i}{\hbar}\vec p\cdot\vec r}\]

A function
\begin{equation}\label{pot:tm:sol}
\Psi(\vec r,t)=
\begin{cases}\psi(\vec r,t), & \mbox{  for $t\leq 0$}\,,\\
\int d^3p\,g(\vec p)\,e^{-i\frac{p^2}{2m\hbar}t}\,e^{\frac{i}{\hbar}\vec p\cdot\vec r}, &
\mbox{ for $t> 0$}\,.
\end{cases}
\end{equation}
is a solution of the Schr\"odinger equation
\[i\hbar\frac{\partial\Psi}{\partial t}=\hat H\Psi\,,\]
and the wave function $\Psi(\vec r,t)$ is continuous at $t=0$.
This continuity condition gives
\begin{equation}\label{cont}
    \psi(\vec r,0)=\int d^3p\,g(\vec p)\,e^{\frac{i}{\hbar}\vec p\cdot\vec r}\,.
\end{equation}
If a function $\psi$ is a stationary solution then
\[\psi_E(\vec r,t)=\phi_E(\vec r)\,e^{-\frac{iE}{\hbar}t}\,,\]
with $\phi_E$ satisfying a stationary Schr\"odinger equation
\[-\frac{\hbar^2}{2m}\Delta\phi_E + U(\vec r)\phi_E=E\phi_E\,.\]
Eq. \eqref{cont} gives now
\begin{equation}\label{cont:stat}
        \phi_E(\vec r)=\int d^3p\,g(\vec p)\,e^{\frac{i}{\hbar}\vec p\cdot\vec r}\,.
\end{equation}

So we have got a general interpretation of the Fourier transform of a wave function. This
gives the momentum distribution of a particle which was influenced by a potential and at
time $t=0$ was suddenly freed. There is no question here about the energy conservation
because of the time-dependency of the hamiltonian \eqref{pot:tm}.

Let us take as an example a well known Biblical story
\subsection{\label{subs:david}How Goliath was defeated by David}
David's sling can be considered as a two-dimensional quantum rotator with a potential
\begin{equation*}
U(\vec r)=\frac{1}{2}m\omega^2(x^2+y^2)\,.
\end{equation*}
Stationary solutions corresponding
to the energy
\begin{equation*}
E_{n_1,n_2}=\hbar\omega(n_1+n_2+1)\,,
\end{equation*}
are given by
\begin{equation}\label{dav1}
  \phi_{n_1,n_2}(x,y)=C_{n_1,n_2}e^{-\frac{m\omega}{2\hbar}(x^2+y^2)}H_{n_1}\left(x\sqrt{\frac{m\omega}{\hbar}}\right)
H_{n_2}\left(y\sqrt{\frac{m\omega}{\hbar}}\right)\,.
\end{equation}
If Goliath were hit directly by a stone still on a cord then he would absorb an impact
energy $E_{n_1,n_2}$. But if a stone was freed from the sling then its momentum
distribution was given by the Fourier transform of the function \eqref{dav1}. So the
probability distribution to have a stone with a momentum between $p$ and $p+dp$ is
proportional to
\begin{equation}
  e^{-\frac{{p_x}^2+{p_y}^2}{m\omega\hbar}}H_{n_1}^2\left(\frac{p_x}{\sqrt{m\omega\hbar}}\right)
  H_{n_2}^2\left(\frac{p_y}{\sqrt{m\omega\hbar}}\right)\,.
\end{equation}
The corresponding impact energy is $p^2/2m$, in general different from $E_{n_1,n_2}$. It
is  easy to check that it is more probable to get the impact energy \emph{lower} than
$E_{n_1,n_2}$. However there is finite, although exponentially decreasing probability,
that a high momentum stone would be thrown. One should notice that such an effect is
impossible for a classical (\emph{i.e.} not-quantum) sling. An exponentially small
probability was not a problem in the considered case taking into account David's
Protector. The crucial point was here a quantum nature of the sling.

\section{\label{sect:math}Mathematical preliminaries}
A cornerstone of quantum mechanics is a precise mathematical interpretation to the notion
of observables. To each observable there corresponds a self-adjoint operator in the
Hilbert state \cite{neumann:mfqm}. For unbounded symmetric operators there was a
nontrivial problem to find all self-adjoint extensions but it was solved long ago
\cite{neumann:slf,dunf,naim}. To give a careful mathematical definition of operators
related to observables it not a matter of a mathematical pedantry. Even in the simplest
case of one dimensional infinite square well a lack of precision leads to obvious
paradoxes \cite{bonneau:2001zq}.

Let us consider a differential operator $-i\frac{d}{dx}$ in the Hilbert space $L_2(0,a)$.
Since:
\begin{equation*}
    \int\limits_0^a dx \bar f\frac{dg}{dx} = \left.\bar f g\right\vert^a_0 -
\int\limits_0^a dx \frac{d\bar f}{dx}g\,,
\end{equation*}
there are infinitely many self-adjoint extensions of the operator $-id/dx$. These
extensions are parameterized by a continuous parameter $\sigma\in[0,2\pi)$ and are
defined on domains
\begin{equation}\label{bound1}
\mathcal{D}_\sigma=\{f\colon f(a)=e^{i\sigma}f(0)\}\,.
\end{equation}
Corresponding eigenvalues are
\begin{equation}\label{eigvl1}
\lambda_n^{(\sigma)}=\frac{\sigma}{a}+\frac{2\pi n}{a}\,,
\end{equation}
and normalized eigenfunctions
\begin{equation}\label{einfn1}
    f_n^{(\sigma)}(x)=
    \begin{cases}
      \frac{1}{\sqrt{a}}\,e^{i\frac{\sigma}{a}x}\,e^{i\frac{2\pi
    n}{a}x}\,,& \mbox{  for $0\leq x\leq a$}\\
    0\,, & \mbox{  for $x$ everywhere else}
    \end{cases}
\end{equation}
where $n=0,\pm 1,\pm 2\dots$

Self-adjoint operators
\begin{equation*}
    \hat p_{(\sigma)}=-i\hbar\frac{d}{dx}\,,
\end{equation*}
defined on the domains $\mathcal{D}_\sigma$ will henceforth be called $\sigma$-momentum
operator. Standard solutions of the infinite potential well take as the ``physical
momentum" the operator $\hat p_{(0)}$ and other extensions are simply rejected. We are
going to show that other $\sigma$-momenta have also well established physical meaning.

To consider the energy operator one should look for a self-adjoint extension of the
operator $d^2/dx^2$. Here the situation is more involved. It was shown \cite{naim1,dunf1}
that domains of self-adjoint extensions are given by a set of boundary conditions
\begin{subequations}\label{bound:h1}
\begin{eqnarray}
  \alpha_{11}f(0)+\beta_{11}f(a)-\alpha_{12}f'(0)-\beta_{12}f'(a) &=& 0 \,,\\
  \alpha_{21}f(0)+\beta_{21}f(a)-\alpha_{22}f'(0)-\beta_{22}f'(a) &=& 0 \,,
\end{eqnarray}
\end{subequations}
with coefficients $\alpha_{ij}$ and $\beta_{kl}$ satisfying
\begin{subequations}\label{bound:h1c}
\begin{align}
  \alpha_{11}\bar\alpha_{12} - \alpha_{12}\bar\alpha_{11} &= \beta_{11}\bar\beta_{12} - \beta_{12}\bar\beta_{11}\,,\\
  \alpha_{21}\bar\alpha_{22} - \alpha_{22}\bar\alpha_{21} &= \beta_{21}\bar\beta_{22} -
  \beta_{22}\bar\beta_{21}\,.
\end{align}
\end{subequations}
In the case of the infinite potential well \eqref{well} a natural choice is to impose on
the wave functions boundary conditions
\begin{subequations}
\begin{equation}\label{bound:well1}
     f(0)=f(a)=0\,,
\end{equation}
which are consistent with the continuity of the wave function. This choice corresponds to
coefficients $\alpha_{ij}$ and $\beta_{kl}$
\begin{eqnarray}\label{bound:well2}
    \alpha_{11}=1\,,&\quad\beta_{11}=-1\,,\quad\alpha_{12}=0\,,\quad\beta_{12}=0\,,\\
    \alpha_{21}=1\,,&\quad\alpha_{22}=0\,,\quad\beta_{21}=0\,,\quad\beta_{22}=0\,.
\end{eqnarray}
This means that functions satisfying boundary conditions \eqref{bound:well1} form a
domain  $\mathcal{D}_\Pi$ of the self-adjoint extension of the operator $d^2/dx^2.$
\end{subequations}
In the case of a particle on a circle a natural choice is to impose on the wave functions
boundary conditions
\begin{subequations}
\begin{equation}\label{bound:well3}
     f(0)=f(a)\,,\quad f'(0)=f'(a)\,.
\end{equation}
This choice corresponds to coefficients $\alpha_{ij}$ and $\beta_{kl}$
\begin{eqnarray}\label{bound:well2al}
    \alpha_{11}=1\,,&\quad\beta_{11}=-1\,,\quad&\alpha_{12}=0\,,\quad\beta_{12}=0\,,\\
    \alpha_{21}=0\,,&\quad\alpha_{22}=1\,,&\beta_{21}=0\,,\quad\beta_{22}=-1\,.
\end{eqnarray}
\end{subequations}
It is remarkable that the intersection of all admissible domains of $\sigma$-momenta is
\begin{equation}\label{inters}
    \bigcap_\sigma\mathcal{D}_\sigma=\{f\colon f(a)=f(b)=0\}=\mathcal{D}_\Pi\,.
\end{equation}
This property makes the extension \eqref{bound:well1} exceptional, at least from the
point of view of momentum operators. A kinetic term $d^2/dx^2$ with this domain is well
defined in (not \emph{on}!) domains of all $\sigma$-momenta.

$\mathcal{D}_\Pi$ is a dense set in the Hilbert space $L_2(0,a)$
as the domain of a self-adjoint operator. This set is too small,
however, to define on it a self-adjoint extension of the operator
$id/dx$. But the property \eqref{inters} together with the general
\cite{naim2}
\begin{thm}
  Any function from the domain of a self-adjoint operator $\mathcal{A}$ can be expanded in
  an uniformly convergent series of eigenfunctions of this operator.
\end{thm}
allows to write
\begin{cor}\label{cor1}
  Any energy eigenfunction \eqref{sol} can be expanded in an uniformly convergent series of
eigenfunctions of any $\sigma$-momentum.
\end{cor}
We have also
\begin{cor}
  $\sigma$-momentum eigenfunctions \eqref{einfn1} cannot be represented as uniformly convergent series of
  energy eigenfunctions.
\end{cor}
Both corollaries can be stated as follows
\begin{quote}
\emph{  In the infinite potential well $\sigma$-momentum representations of stationary
  states are always uniformly convergent. Energy representations of $\sigma$-momentum
  eigenfunctions are not uniformly convergent.}
\end{quote}
Let us make a mathematical exercise to calculate

\subsection{\label{subs:sgm:repr}$\sigma$-momentum representation of stationary states}
We have
\begin{equation}\label{exp:en}
    \sqrt{\frac{2}{a}}\sin{\frac{N\pi}{a}x} =
e^{i\frac{\sigma}{a}x}\frac{1}{\sqrt{a}}\sum\limits_{n=-\infty}^{+\infty}c_n(\sigma)
e^{i\frac{2\pi n}{a}x}\,.
\end{equation}
Coefficients $c_n(\sigma)$ are given here as
\begin{equation}\label{exp:en:coe}
\frac{\sqrt{2}}{a}\int\limits_0^a
    dx\,e^{-i\left(\frac{\sigma}{a}+\frac{2\pi n}{a}\right)x}\sin{\frac{\pi N}{a}x}\ =
    \pi N\sqrt{2}\frac{e^{-i\sigma}(-1)^N -1}{(\sigma+2\pi n)^2-\pi^2 N^2}\,.
\end{equation}
It is convenient to discuss cases of even and odd $N$ separately.

If $N=2r$ we can write
\begin{equation}\label{coe-even1}
    c_n(\sigma)=
    -4i\pi r\sqrt{2}e^{-i\frac{\sigma}{2}}\frac{\sin{\frac{\sigma}{2}}}{(\sigma+2\pi n)^2- 4\pi^2 r^2}\,,
\end{equation}
A special care is needed when the nominator of this expression is equal to zero. For
$\sigma=0$ one gets then
\begin{equation}\label{coe:even2}
    c_n(0)=\frac{1}{i\sqrt{2}}
    \begin{cases}
      \ 1, &\text{for } n=r\,,\\
      -1, &\text{for } n=-r\,,\\
      \ 0, &\text{in other cases}\,.
    \end{cases}
\end{equation}
After substitution to Eq. \eqref{exp:en} this gives a consistency check
\begin{equation}\label{en:even:expn}
    \psi_{2r}(x) = \frac{1}{2i}\sqrt{\frac{2}{a}}\left(e^{i\frac{2\pi
r }{a}x}-e^{-i\frac{2\pi r}{a}x}\right)\,.
\end{equation}
For nonzero  $\sigma$ we have
\begin{equation}\label{exp:ev}
    \psi_{2r}(x)= -4\pi i r \sqrt{\frac{2}{a}}\,e^{-i\frac{\sigma}{2}}\,
    e^{i\frac{\sigma}{a}x}\sin{\frac{\sigma}{2}}\sum\limits_{n=-\infty}^{+\infty}
    \frac{e^{i\frac{2\pi n}{a}x}}{(\sigma+2\pi n)^2-4\pi^2 r^2}\,.
\end{equation}
 If $N=2r+1$ we can write Eq. \eqref{exp:en:coe} as
\begin{equation}\label{coe:odd1}
       c_n(\sigma)=
    -2\pi(2r+1)\sqrt{2}e^{-i\frac{\sigma}{2}}\frac{\cos{\frac{\sigma}{2}}}
    {(\sigma+2\pi n)^2-\pi^2(2r+1)^2}\,.
\end{equation}
For $\sigma=\pi$ one gets similarly like in Eqs \eqref{coe:even2}
\begin{equation}\label{coe:odd2}
    c_n(\pi)=\frac{1}{i\sqrt{2}}
    \begin{cases}
      \ 1, &\text{for } n=r\,,\\
      -1, &\text{for } n=-r-1\,,\\
      \ 0, &\text{in other cases}\,.
    \end{cases}
\end{equation}
This gives, similarly like in Eq. \eqref{en:even:expn},
\begin{equation}\label{en:odd:expn}
        \psi_{2r+1}(x)=\frac{1}{2i}\sqrt{\frac{2}{a}}\left(e^{i\frac{2(r+1)\pi
}{a}x}-e^{-i\frac{2(r+1)\pi}{a}x}\right)\,.
\end{equation}
For $\sigma\neq\pi$ we have
\begin{equation}\label{exp:odd}
    \psi_{2r+1}(x)= -2\pi(2r+1)\sqrt{\frac{2}{a}}\,e^{-i\frac{\sigma}{2}}\,
    e^{i\frac{\sigma}{a}x}\cos{\frac{\sigma}{2}}\sum\limits_{n=-\infty}^{+\infty}
    \frac{e^{i\frac{2\pi n}{a}x}}{(\sigma+2\pi n)^2-\pi^2 (2r+1)^2}
\end{equation}
All these mathematical expansions from Eqs \eqref{en:even:expn},
\eqref{exp:ev}, \eqref{en:odd:expn}, and \eqref{exp:odd} would
have physical meaning with a satisfactory physical interpretation
of $\sigma$-momenta. This will be done in the next section. It
should be now noted that the choice $\sigma=0$ gives expansions of
the potential well stationary states into momentum eigenfunctions.
The momentum spectrum is given then by Eq. \eqref{eigvl1} with
$\sigma=0$. An elusively simple equation \eqref{spr} \emph{is not
always} a momentum expansion because $N\pi/a$ are allowed momenta
only for even $N$. Only in such cases a stationary state can be
visualized as the superposition of two waves with opposite
momenta. For odd $N$ the momentum expansion is
\begin{equation}\label{exp:odd:zr}
    \psi_{2r+1}(x)= -2\pi(2r+1)\sqrt{\frac{2}{a}}\,\sum\limits_{n=-\infty}^{+\infty}
    \frac{e^{i\frac{2\pi n}{a}x}}{4\pi^2 n^2-\pi^2 (2r+1)^2}\,,
\end{equation}
with a much richer structure.

More detailed analysis of this problem will be performed in
Section \ref{subs:mom:mv:well}.
\section{\label{sect:mom:mv}Momentum seen from the moving reference frame}

We are going to find physical meaning of different self-adjoint extensions of the
operator $-id/dx$. It is a standard procedure to identify this operator with the
translation generator. It is not enough, however, to relate this to the physical
momentum. The same differential operator can be also related to a component of the
angular momentum even for the same boundary conditions.

Let us consider as an example the operator $-i\hbar d/dx$ in the Hilbert space
$L_2(0,2\pi)$ defined on the domain
\begin{equation}\label{bound2pi}
    \mathcal{D}_0=\{f\colon f(2\pi)=f(0)\}\,.
\end{equation}
A spectrum of this operator
\begin{equation}\label{eigvl2pi}
\lambda_n^{(0)}=n\hbar\,,
\end{equation}
and normalized eigenfunctions
\begin{equation}\label{einfn2pi}
    f_n^{(0)}(x)=\frac{1}{\sqrt{2\pi}}\,e^{i n x}\,,
\end{equation}
are the same both for the momentum in the interval $(0,2\pi)$ as for the third component
of the angular momentum when the variable $x\in(0,2\pi)$ is interpreted as an angular
variable.

To get a momentum operator proper transformation properties are needed, specific for the
corresponding classical variable. Let us consider two coordinate systems
$\mathcal{O}(x,t)$ and $\mathcal{O'}(\zeta,\tau)$ related by the Galilei transformation
\begin{equation}\label{transf}
 x=\zeta+ V\tau\,;\qquad t=\tau\ .
\end{equation}
The following discussion is based on the ``passive point of view" when the same system is
observed by different observers $A$ in $\mathcal{O}$ and $A'$ in $\mathcal{O}'$ having
different relations to the system.

Let $-i\hbar d/dx$ be a momentum operator in the system $\mathcal{O}$ and let $f_\lambda$
be an eigenfunction of the momentum operator associated with the eigenvalue $\lambda$.
The momentum operator should fulfill the following conditions:
\begin{subequations}\label{mom:mv:cond}
\begin{itemize}
    \item a momentum operator has the same structure in all inertial systems \emph{i.e.}
    \begin{equation}\label{mom:mv:cond1}
      -i\hbar d/d\zeta \mbox{ is a momentum operator in the coordinate system }\mathcal{O}',
    \end{equation}
    \item a physical state with a defined momentum in one inertial system has a definite
    momentum in any inertial system \emph{i.e.}
    \begin{equation}\label{mom:mv:cond2}
    f_\lambda\mbox{ is transformed into  }
        \tilde f_{\tilde\lambda}\colon\quad -i\hbar\frac{d\,\tilde f_{\tilde\lambda}}{d\zeta} =
    \tilde\lambda\tilde f_{\tilde\lambda}\,,
    \end{equation}
    \item eigenvalues of the momentum operator transform under the Galilei transformation
    like their classical counterparts \emph{i.e.}
    \begin{equation}\label{mom:mv:cond3}
    \tilde\lambda=\lambda - mV\,.
    \end{equation}
\end{itemize}
\end{subequations}

We make an ansatz
\begin{equation}\label{anstz1}
\tilde f_{\tilde\lambda}(\zeta,\tau)=e^{ig(\zeta,\tau)}f(\zeta+V\tau)\,.
\end{equation}
This gives
\begin{equation}\label{transf1}
\begin{split}
    -i\hbar\frac{d\tilde f}{d\zeta} &= \hbar\frac{d g}{d\zeta}e^{i g} f -
    i\hbar e^{ig}\frac{df_\lambda}{dx} \\
 &= \hbar\frac{d g}{d\zeta}\tilde f + \lambda\tilde f = \tilde\lambda\tilde f\,.
\end{split}
\end{equation}
The correspondence rule \eqref{mom:mv:cond3} gives
\begin{equation}\label{transf2}
    \hbar\frac{d g}{d\zeta} + \lambda = \lambda - m V\,.
\end{equation}
A general solution has a form
\begin{equation}\label{expfactor}
    g(\zeta,\tau)=-\frac{m V}{\hbar}\zeta + T(\tau)\,,
\end{equation}
where $T$ is an arbitrary function of the variable $\tau$.

Starting from the consistency conditions \eqref{mom:mv:cond} we have obtained a general
transformation rule for momentum eigenfunctions under the Galilei transformation
\begin{equation}\label{trnsf:gen:eig}
    \tilde f_{\lambda-mV}(\zeta,\tau) = e^{-i\left(\frac{m V}{\hbar}\zeta -
    T(\tau)\right)}\,f_\lambda(\zeta+V\tau)\,.
\end{equation}
Because of properties of self-adjoint operators this rule gives transformation rules for
any vector from the Hilbert space.

If the momentum operator has a point spectrum, then its eigenfunctions form a base in a
Hilbert space. This is a case for the $L_2(0,a)$ space. Any element $u$ of this space can
be expanded as
\begin{equation}\label{exp:gen1}
    u(x)=\sum\limits_\lambda c_n\tilde f_\lambda(x)\,.
\end{equation}
It follows from this that an observer from the Galilei transformed reference frame
\eqref{transf} sees this vector as
\begin{equation}\label{exp:gen2}
    \tilde u(\zeta,\tau)= e^{-i\left(\frac{m V}{\hbar}\zeta -
    T(\tau)\right)}\sum\limits_\lambda c_n f_\lambda(\zeta+V\tau)=e^{-i\left(\frac{m V}{\hbar}\zeta -
    T(\tau)\right)}u(\zeta+V\tau)\,.
\end{equation}

A function $T$ is fixed by subsidiary conditions fulfilled by the function $u$. It will
be shown in Section \ref{sect:schr:mv} that for one particle Schr\"odinger equation a
function $T(\tau)$ has a form $-mV^2\tau/2$.

A result \eqref{exp:gen2} can be easily generalized to the case of a continuous spectrum
of the momentum operator. That is a standard mathematical procedure
\cite{naim,dunf,riesz} equivalent to the replacement of the sum in Eq. \eqref{exp:gen1}
by the Fourier integral.

For the moment we restrict ourselves to
\subsection{\label{subs:mom:var:well}Momentum inside the infinite potential well}
The momentum observable ``at rest" is $-i\hbar d/dx$ with the domain
\begin{equation}\label{bound:st}
\mathcal{D}_0=\{f\colon f(a)=f(0)\}\,.
\end{equation}
Boundary conditions in the moving reference frame $\mathcal{O'}(\zeta,\tau)$ are given at
points $\zeta=-V\tau$ and $\zeta=-V\tau + a$. Using transformations rules given by Eqs
\eqref{transf1} and \eqref{expfactor} one gets the boundary conditions for the function
$\tilde f$
\begin{equation}\label{boundmv}
    \tilde f(-V\tau+a,\tau)=e^{-i\frac{mV}{\hbar}a}\tilde f(-V\tau,\tau)\,.
\end{equation}
This gives us a direct interpretation of $\sigma$-momenta:

\emph{\begin{quote}
$\sigma$-momentum it is the momentum observable measured by the
observer moving with a velocity $V$ such that
\begin{equation}\label{mommv}
    \sigma=-\frac{m V}{\hbar}a\mod 2\pi\,.
\end{equation}
\end{quote}}
%%%%%%%%%%%%%%%%%%%%%%%%%%%%%%%%%%%%%%%%%%%%%%%%%%%%%%%%%%%%%%%%%%%%%%%%%%%%%%%
To make the $\sigma$-momentum a real quantum mechanical momentum one should check the
validity of Canonical Commutation Relations (CCR) of a momentum and a position operator
in our system. In general, a problem of CCR on the finite interval is far from being
obvious \cite{Lassner:1987}. As opposed to the entire real line case where both position
$\hat{X}$ and momentum $\hat{P}$ operators are unbounded, here the operator $\hat{X}$ is
bounded in the Hilbert space $L_2(0,a)$ whereas the operator $\hat{P} = -i\hbar\partial$
is unbounded. This leads to technical troubles related  to the domain
$\mathcal{D}([\hat{X},\hat{P}])$ of the commutator $[\hat{X},\hat{P}]$. The domain where
CCR are fulfilled is
\begin{equation}\label{domainCCR1}
    \mathcal{D}(\hat{P}\hat{X})\cap\mathcal{D}(\hat{X}\hat{P})\,,
\end{equation}
where $\mathcal{D}(\hat{P}\hat{X})$ and $\mathcal{D}(\hat{X}\hat{P})$ are domains of
operator products $\hat{P}\hat{X}$ and $\hat{X}\hat{P}$ correspondingly.

The domain $\mathcal{D}(\hat{P}\hat{X})$ is
\begin{equation}\label{domainXP}
    \mathcal{D}(\hat{P}\hat{X}) = \{f\colon (\hat{X}f)\in\mathcal{D}(\hat{P})\}\,,
\end{equation}
and the domain of the product $\hat{X}\hat{P}$ is equal here to the domain of the
momentum operator because of the boundedness of the position operator.

 For the $\sigma$-momentum $\hat{p}_{(\sigma)}$ the domain $\mathcal{D}_\sigma$ is
given by \eqref{bound1}. Then
\begin{equation}\label{domainCCR2}
    \mathcal{D}([\hat{X},\hat{p}_{(\sigma)}]) = \{f\colon
(\hat{X}f)(a)=e^{i\sigma}(\hat{X}f)(0)\}\cap\{f\colon f(a)=e^{i\sigma}f(0)\} = \{f\colon
f(a)=f(b)=0\}\,.
\end{equation}
So CCR are realized on the dense domain in $L_2(0,a)$. This domain does not depend on the
$\sigma$-realization of the momentum operator and coincides with the domain $D_\Pi$
\eqref{inters} of the energy operator.

Different $\sigma$-momenta, as corresponding to unitary
non-equivalent projective representation of the Galilei group,
correspond to different unitary non-equivalent representations of
CCR although all are realized on the same dense domain $D_\Pi$.

%%%%%%%%%%%%%%%%%%%%%%%%%%%%%%%%%%%%%%%%%%%%%%%%%%%%%%%%%%%%%%%%%%%%%%%%%%%%%%%
Coming back to our quantum squash model: a player running with the velocity $V$ sees a
squash ball having $\sigma$-momentum. $\sigma$ is given here by Eq. \eqref{mommv}.
H(is)er momentum eigenfunctions take on the form
\begin{equation}\label{egnfn:mom:mv}
    \tilde f_n(\zeta,\tau)=e^{-i\frac{mV}{\hbar}\zeta}\,e^{i\frac{2\pi
    n}{a}(\zeta+V\tau)}= e^{\frac{i}{\hbar}(\frac{2\pi n}{a} - mV)\zeta}\, \,e^{i\frac{2\pi
    n}{a}V\tau}\,.
\end{equation}

It is also interesting to look for solutions of the infinite potential well observed by a
running player. This will be the subject of the next section.

\section{\label{sect:schr:mv}Schr\"odinger equation seen from the moving reference frame}
Let us consider a particle subjected to the influence of a time-dependent potential $U$.
In the coordinate system $\mathcal{O}(x,t)$ the Schr\"odinger equation takes on the form
\begin{equation}\label{schr1}
    i\hbar\frac{\partial\Psi}{\partial t}=-\frac{\hbar^2}{2m}\frac{d^2\Psi}{\partial
    x^2}+U(x,t)\Psi\,.
\end{equation}
This equation is obviously not Galilei-invariant unless the potential is a trivial
constant. A typical procedure is to investigate physical consequences of the symmetry
group starting from the symmetry-invariant equations. In the case of the Galilei (or
Poincar\'e) group this leads to a free particle wave function realizing a unitary
representation of the group. For the Galilei group and the Schr\"odinger equation one
gets \cite{Inonu:1952ct,bargm,hamerm,levy:1963} that only nontrivial projective
representations are physical realizations of the symmetry.

We are going to consider a more general approach based on the equivalence of all inertial
coordinate systems. This Galilean equivalence principle demands that all laws of physics
take the same form in different frames connected by the Galilei (or Poincar\'e for
relativistic theory) transformations. We derive from the postulates that ``the Galilei
transformation is true'' and ``the Schr\"odinger equation is true'' the transformation
law of wave function for any scalar potential. So we do, what Galilei would do, ``if
Galilei had know quantum mechanics" \cite{kaem}.

An observer in the reference frame $\mathcal{O}'(\zeta,\tau)$ sees the potential $U$ as
\begin{equation}\label{pot_mv}
    \widetilde{ U}(\zeta,\tau)=U(\zeta+V\tau,\tau)\,.
\end{equation}
It is assumed here that the potential is a scalar with respect to the Galilei
transformation \eqref{transf}. The equivalence principle demands that a wave function
$\widetilde{\Psi}(\zeta,\tau)$ viewed by an observer $A'$ in the coordinate system
$\mathcal{O}'$ satisfies the Schr\"odinger equation
\begin{equation}\label{schr2}
    i\hbar\frac{\partial\widetilde{\Psi}}{\partial\tau} =
    -\frac{\hbar^2}{2m}\frac{d^2\widetilde{\Psi}}{\partial\zeta^2}+
    \widetilde{U}(\zeta,\tau)\Psi\,.
\end{equation}
An ansatz
\begin{equation}\label{subst}
    \widetilde{\Psi}(\zeta,\tau) = e^{i u(\zeta,\tau)}\Psi(\zeta+V\tau,t)\,,
\end{equation}
gives
\begin{align*}
  \frac{\partial\widetilde{\Psi}}{\partial\tau} &= i e^{i u}\frac{\partial u}{\partial\tau}\Psi +
e^{i u}\frac{\partial\Psi}{\partial x}V + e^{i u}\frac{\partial\Psi}{\partial\tau}\,, \\
  \frac{\partial\widetilde{\Psi}}{\partial\zeta} &= i e^{i u}\frac{\partial u}{\partial\zeta}\Psi +
e^{i u}\frac{\partial\Psi}{\partial x}\,, \\
  \frac{\partial^2\widetilde{\Psi}}{\partial\zeta^2} &= i e^{i u}\frac{\partial^2 u}{\partial\zeta^2}\Psi -
e^{i u}\left(\frac{\partial u}{\partial\zeta}\right)^2\Psi + 2 i e^{i u}\frac{\partial
u}{\partial\zeta}\frac{\partial\Psi}{\partial x} + e^{i u}\frac{\partial^2\Psi}{\partial
x^2}\,.
\end{align*}
We see that Eq. \eqref{schr2} is fulfilled if
\begin{subequations}\label{consxx}
  \begin{equation}\label{cons1}
    i\hbar\frac{\partial\Psi}{\partial x}V=-\frac{\hbar^2}{2 m}2 i\frac{\partial u}{\partial\zeta}\frac{\partial\Psi}{\partial x}\,,
\end{equation}
and
\begin{equation}\label{cons2}
    -\hbar\frac{\partial u}{\partial\tau}=\frac{\hbar^2}{2
    m}\frac{m^2}{\hbar^2}V^2\,.
\end{equation}
\end{subequations}
A solution of Eqs \eqref{consxx} takes on the form
\begin{equation}\label{phase1}
    u(\zeta,\tau)=-\frac{m}{\hbar}V\zeta - \frac{mV^2}{2\hbar}\tau + C(V)\,.
\end{equation}
So wave functions in different inertial reference frames connected by the Galilei
transformation \eqref{transf} are connected (up to the constant phase factor $e^{iC}$) by
the relation
\begin{equation}\label{phase}
    \widetilde{\Psi}(\zeta,\tau) = e^{-\frac{i}{\hbar}\left(mV\zeta +
    \frac{mV^2}{2}\tau\right)}\Psi(x,t)\,.
\end{equation}
We have got the same factor as obtained by Bargmann \cite{bargm} for a free Schr\"odinger
particle subjected to the Galilei transformation. This factor leads to the
mass-superselection rule what is mathematically due to the fact that projective (ray)
representations of the Galilei group are not unitary equivalent to the usual
representations \cite{hamerm}.

\subsection{Stationary states seen from the moving reference frame}

Let us consider now a particle in a static potential $U(x)$ described by a stationary
wave function
\begin{align}
    \Psi_{n}(x,t) & = \psi_{n}(x)e^{-\frac{i}{\hbar}E_n t}\,,\label{stat_wf1}\\
    -\frac{\hbar^2}{2m}\frac{d^2\psi_{n}}{dx^2}+U(x)\psi_{n}(x)& =  E_n\psi_{n}(x)\,.\label{stat:wf2}
\end{align}
According to the general rule \eqref{phase}, this state when
viewed by a moving observer from the reference frame
$\mathcal{O'}(\zeta,\tau)$ is described by a wave function
\begin{equation}\label{stat:wf1:mv}
    \widetilde{\Psi}_n(\zeta,\tau)=e^{-\frac{i}{\hbar}m V \zeta}\psi_n(\zeta +
    V\tau)e^{-\frac{i}{\hbar} \left(E_n+\frac{m V^2}{2}\right)\tau}\ .
\end{equation}
One should note that this is not an energy eigenstate. This follows, at least formally,
from the fact that Galilei transformed potential $\tilde U(\zeta,\tau)$ is now time
dependent. The energy, however, is still conserved. To check this let us calculate an
average value of the energy for the state described by the wave function
\eqref{stat:wf1:mv}. It is given as
\begin{equation}\label{energ:av1}
    \langle E\rangle_n = i\hbar\int
    d\zeta\widetilde{\Psi}_n^*(\zeta,\tau)\frac{\partial}{\partial\tau}\widetilde{\Psi}_n(\zeta,\tau)\ .
\end{equation}
Taking into account that solutions of Eq. \eqref{stat:wf2} are real one obtains then
\begin{equation}\label{energ:av2}
\langle E\rangle_n = E_n + \frac{m V^2}{2}\,.
\end{equation}
This result seems to be surprising even in the simplest case of a free particle with the
momentum $p$. Taking into account that the energy $E=p^2/2m$ and the momentum is Galilei
transformed to $p-mV$ one should expect in Eq. \eqref{energ:av2} a subsidiary term of the
form $-pV$. However, this not the case. Energy $p^2/2m$ is ``produced" by a particle with
the momentum $\pm p$. A real stationary state is a superposition of two waves
corresponding to opposite momenta $\pm p$. They are Galilei transformed to $p\mp mV$
correspondingly and give contributions to the energy $p^2/2m + mV^2/2 \mp pV.$ These are
additive contributions to the total energy so terms $\pm pv$ cancel each other. This
remark gives a perfect agreement of Eq. \eqref{energ:av2} with our classical intuition
although, as we'll see later, does not always agree with a quantum reality.

\subsection{\label{subs:move well}Infinite potential well seen from the moving reference
frame}

When the  potential well \eqref{well} is observed from the moving reference frame
$\mathcal{O}'$ it is seen as
\begin{equation}\label{potwell:mv}
    U(\zeta + V\tau)=
    \begin{cases}
      0\,;\quad & \text{if } - V\tau\leq\zeta\leq a - V\tau\,,\\
      \infty \,;\quad & \text{if } \zeta\notin \langle - V\tau,a - V\tau\rangle\,.
    \end{cases}
\end{equation}

The wave function satisfies boundary conditions
\begin{equation}\label{bound:mv}
   \widetilde{\Psi}(- V\tau,\tau)= \widetilde{\Psi}(a- V\tau,\tau)=0\ .
\end{equation}
The solution \eqref{stat:wf1:mv} has now a form
\begin{equation}\label{well:mv}
    \widetilde{\Psi}_N(\zeta,\tau)=
    \begin{cases}
    \sqrt{\frac{2}{a}}e^{-\frac{i}{\hbar}m V \zeta}\sin{\frac{N\pi}{a}(\zeta+V\tau)}
    e^{-\frac{i}{\hbar}\left(\frac{\pi^2\hbar^2}{2 m a^2}N^2\ + \frac{m V^2}{2}\right)\tau}
    \,;\ & \text{if }  - V\tau\leq\zeta\leq a - V\tau\,,\\
    0\,;\ & \text{if } \zeta\notin \langle - V\tau,a - V\tau\rangle\,.
    \end{cases}
\end{equation}
where $N=1,2,\dots$

This solution can be written in the region $- V\tau\leq\zeta\leq a - V\tau$ as a
superposition of two plain waves
\begin{subequations}\label{sol:well}
\begin{equation}
  \widetilde\Psi_N(\zeta,\tau)=\psi_{(+),N}(\zeta,\tau)-\psi_{(-),N}(\zeta,\tau)
\end{equation}
where
\begin{align}
  \psi_{(+),N}(\zeta,\tau) & =  \frac{1}{2i}\sqrt\frac{2}{a}\,e^{\frac{i}{\hbar}\left(\frac{N\pi\hbar}{a}-m
  V\right)\zeta}\,
e^{-\frac{i}{2m\hbar}\left(\frac{N\pi\hbar}{a}-m V\right)^2\tau}\,,\\
    \psi_{(-),N}(\zeta,\tau)& = \frac{1}{2i}\sqrt\frac{2}{a}\,e^{-\frac{i}{\hbar}\left(\frac{N\pi\hbar}{a}+ m
    V\right)\zeta}\,
e^{-\frac{i}{2m\hbar}\left(\frac{N\pi\hbar}{a}+ m V\right)^2\tau}\,.
\end{align}
\end{subequations}
This decomposition confirms our semiclassical understanding of the quantum problem. Alas,
our semiclassical understanding does not quite agree with the mathematics behind the
scene.

\subsection{\label{subs:mom:mv:well}Moving observer measures momentum in the well}

A measurement of an observable is mathematically equivalent to the
spectral decomposition of the wave function into corresponding
eigenfunctions. A general mathematical decomposition into momentum
eigenfunctions was done in Section \ref{subs:sgm:repr}. A physical
problem: ``what are possible momenta measured by a moving observer
in the infinite potential well?'', will be solved when h(is)er
wave function $\widetilde{\Psi}_N(\zeta,\tau)$ \eqref{well:mv}
will be expanded into h(is)er momentum eigenfunctions $\tilde
f_n(\zeta,\tau)$ \eqref{egnfn:mom:mv}
\begin{equation}\label{expns:mom:mv}
    \widetilde\Psi_N(\zeta,\tau) = \sum\limits_{n=-\infty}^{+\infty}c_n^{(N)}(\tau)\,\tilde
f_n(\zeta,\tau)\,.
\end{equation}
Coefficients $c_n^{(N)}$ are calculated as
\begin{equation}\label{coeff:mom:mv}
\begin{split}
    c_n^{(N)}(\tau) &= \frac{\sqrt{2}}{a}
    e^{-\frac{i}{\hbar}\left(\frac{\pi^2\hbar^2}{2 m a^2}N^2\ + \frac{m V^2}{2}\right)\tau}
    \,\int\limits_{-V\tau}^{-V\tau+a}d\zeta\,
    \sin{\frac{N\pi}{a}(\zeta + V\tau)}\,e^{-i\frac{2\pi n}{a}(\zeta+V\tau)}\\
    &=e^{-\frac{i}{\hbar}\left(\frac{\pi^2\hbar^2}{2 m a^2}N^2\ + \frac{m
    V^2}{2}\right)\tau}\times
    \begin{cases}
      -\frac{2 N\sqrt{2}}{(4n^2-N^2)\pi}\,,&\mbox{  for odd }N\,,\\
    \pm\frac{1}{i\sqrt{2}}\,,&\mbox{  for even }N,\ n=\pm N/2\,,\\
    0\,,&\mbox{  for even }N,\ n\neq\pm N/2\,.
    \end{cases}
\end{split}
\end{equation}
This gives \underline{for even $N$}  a simple expression, consistent with a semiclassical
approach
\begin{equation}\label{mom:mv:ev}
\widetilde\Psi_N(\zeta,\tau)
    =\frac{1}{2i}\sqrt{\frac{2}{a}}\left[e^{\frac{i}{\hbar }\left(\frac{\pi N\hbar}{a} - mV\right)\zeta}\,
e^{-\frac{i m}{2\hbar}\left(V - \frac{\pi\hbar N}{ma}\right)^2\tau}- e^{-\frac{i}{\hbar
}\left(\frac{\pi N\hbar}{a} + mV\right)\zeta}\, e^{-\frac{i
m}{2\hbar}\left(\frac{\pi\hbar N}{ma} + V \right)^2\tau}\right]
\end{equation}
The case of \underline{for odd $N$} is more involved and the momentum expansion
\eqref{expns:mom:mv} takes on the form
\begin{subequations}
\begin{equation}\label{mom:mv:odd}
\begin{split}
\widetilde\Psi_N(\zeta,\tau)=& \frac{2}{\pi N}\sqrt{\frac{2}{a}}\,
    e^{-i\frac{mV}{\hbar}\zeta}\,e^{-\frac{i}{\hbar}\left(\frac{\pi^2\hbar^2}{2 m a^2}N^2\ +
    \frac{m V^2}{2}\right)\tau}\\
    & - \frac{2N}{\pi}\sqrt{\frac{2}{a}}e^{-i\frac{mV}{\hbar}\zeta}\,
    e^{-\frac{i}{\hbar}\left(\frac{\pi^2\hbar^2}{2 m a^2}N^2\ +
    \frac{m V^2}{2}\right)\tau}\sum\limits_{n=1}^{+\infty}\frac{1}{4n^2-N^2}\left[e^{i\frac{2\pi
    n}{a}(\zeta+V\tau)} + e^{-i\frac{2\pi
    n}{a}(\zeta+V\tau)}\right]\,,
\end{split}
\end{equation}
This can be also written as
\begin{equation}\label{mom:mv:odd2}
\widetilde\Psi_N(\zeta,\tau)
 = \frac{4N}{\pi}\sqrt{\frac{2}{a}}\sum\limits_{n=-\infty}^{+\infty}
\frac{e^{-i\frac{\pi^2\hbar}{2ma^2}(N^2-4n^2)\tau}}{N^2-4n^2}\, e^{\frac{i}{\hbar
}\left(\frac{2\pi\hbar n}{a} - mV\right)\zeta}\, e^{-\frac{i
m}{2\hbar}\left(\frac{2\pi\hbar n}{ma} - V \right)^2\tau}\,,
\end{equation}
\end{subequations}
where contributions from stationary plane waves states  are explicitly selected. We shall
call from this time on a ``stationary plane wave state'' a plain wave with an explicit
time dependence, \emph{i.e.} a function of the form
\begin{equation}\label{pln:wv:st}
    e^{-i\frac{p^2}{2m\hbar}t}\,e^{\frac{i}{\hbar}\vec p\cdot\vec r}\,.
\end{equation}
There is a striking difference in the behavior of odd- and even-$N$ states
$\widetilde\Psi_N$. Any even $N$ state is a superposition of two stationary plane waves
states, while an odd-$N$ state cannot be represented as a superposition of stationary
plane wave states. An underlying mechanism is the same which made difference between Eqs
\eqref{en:even:expn} and \eqref{exp:odd:zr} --- allowed momenta in the infinite well are
not always the same as formal arguments of the energy eigenfunctions \eqref{sol}.

This can be changed with a change of boundary conditions \eqref{bound}. If they are
replaced by periodic-type conditions \eqref{bound:well3}
\begin{equation}\label{bound:perd}
    \psi(0)=\psi(a)\,,\qquad \psi'(0)=\psi'(a)\,,
\end{equation}
then eigenfunctions of the Hamiltonian are
\begin{equation}\label{sol:perd}
\sqrt{\frac{2}{a}}\sin\frac{2N\pi}{a}x\,,\qquad \sqrt{\frac{2}{a}}\cos\frac{2N\pi}{a}x\,.
\end{equation}
We see that for such boundary conditions allowed momenta in the infinite well are always
the same as formal arguments of energy eigenfunctions and any state \eqref{sol:perd} is a
superposition of two plane waves with the definite momenta. So, in a sense, a situation
of periodic boundary conditions is ``better" than for boundaries \eqref{sol}.

It is easy to create a "worse" situation. To this end it is enough to take
antiperiodic-type boundary conditions
\begin{equation}\label{bound:aperd}
    \psi(0)=-\psi(a)\,,\qquad \psi'(0)=-\psi'(a)\,.
\end{equation}
Then eigenfunctions of the Hamiltonian are
\begin{equation}\label{sol:aperd}
\sqrt{\frac{2}{a}}\sin\frac{(2N+1)\pi}{a}x\,,\qquad
\sqrt{\frac{2}{a}}\cos\frac{(2N+1)\pi}{a}x\,.
\end{equation}
For such boundary conditions allowed momenta in the infinite well are never the same as
formal arguments of energy eigenfunctions and there is no state \eqref{sol:aperd} as a
superposition of two plane waves with the definite momenta.

\section{\label{sect:concl}Conclusions}
We performed a careful analysis of the notion of an observable related to the physical
momentum. For the beginning one should identify a self-adjoin operator connected to this
notion. In quantum mechanics defined on a finite interval a situation is more complicated
than in the case of an unrestricted theory on $\mathbb{R}$ because there is a continuum
\eqref{bound1} of self-adjoint extensions of the differential operator $id/dx$.

If you want to go beyond an argument that $-i\hbar d/dx$ is a physical momentum when it
is assigned to the letter ``$p$'', and it is an angular momentum when assigned to the
sign ``$l_z$'', then appropriate transformation properties must be taken into account. In
a similar manner, three numbers can be a finite three-elements set or components of three
dimensional vector. A choice depends on assumed transformation properties with respect to
rotations. In the momentum case, transformation properties are given (in a
nonrelativistic approach) by the Galilei group as was done by conditions
\eqref{mom:mv:cond}. All this together, supplemented with the equivalence of all inertial
coordinate systems, led to different realizations of the momentum operator in the finite
volume ---  in Section \ref{subs:mom:mv:well}. Those different realizations have
different spectra. This is quite obvious from the physical point of view.

Such an approach, based on a generalized correspondence principle \eqref{mom:mv:cond}
gave a physical interpretation of all self-adjoint extensions of the operator $-i\hbar
d/dx$. It was also shown that those different extensions realize different non-unitary
equivalent representations of CCR on the universal dense domain.

Obtained results can be generalized for a three dimensional rectangular box and the
momentum operator  $-i\hbar\nabla$.

Results of Section \ref{sect:schr:mv} show that important physical properties, related to
transformation laws of wave functions, can be obtained under much weaker assumption than
it was done in the past. A condition of Galilei invariance, widely used to obtain mass
superselection rule, is replaced by (generalized) Galilei equivalence principle. This
allows to go beyond a free particle theory and gives results also for an arbitrary scalar
potential.

The transformation law \eqref{stat:wf1:mv} can be treated as a realization of different
self-adjoint extensions of the hamiltonian. This is clearly visible in the infinite
potential well where Eqs \eqref{bound:h1} and \eqref{bound:h1c} give different
self-adjoint extensions of the operator $d^2/dx^2$. However, a situation here is not such
simple as in the case of momentum operators. A structure of self-adjoint extensions of
the operator $d^2/dx^2$ is much richer than in the case of the momentum operator. Only a
part of those extensions can be related to the Galilei transformations and these are done
by Eq. \eqref{well:mv}.

Results related to the momentum distribution, obtained in Sections
\ref{sect:momdstr} and \ref{subs:mom:mv:well}, need some comments.
We have shown that the Fourier integral of the stationary wave
function is directly related to the time-dependent dynamics given
by Eq. \eqref{pot:tm}. This gives a direct interpretation of that,
what is usually called ``momentum representation of the wave
function'' or ``wave function in momentum space''. This
interpretation is different from that what is usually found in
textbooks. A simple statement that a wave function in momentum
space
\begin{equation}\label{wv:fn:mom}
    \Phi(\vec p,t)=\int d^3r\, \Psi(\vec r,t)\,e^{-\frac{i}{\hbar}\vec p\cdot\vec r}\,,
\end{equation}
is a probability amplitude to measure the momentum $\vec p$ at time $t$ \emph{is simply
not true!} It sometimes reasonable, for technical reasons, to use a momentum
representation of the wave function but only because of its mathematical equivalence to
the wave function.

There is an exception in the ``finite Universe'' with dynamics defined on a finite
interval. Let us introduce here two notions of momentum representation. The first is
analogous to the previous one. You take the Fourier integral and you obtain a dynamics as
in Eq. \eqref{pot:tm}. This means that at $t=0$ all impenetrable walls vanish and you are
left with a free particle. Such a situation is used in the HBT effect \cite{hbt56},
originally invented to determine the dimensions of distant astronomical objects. This
method is widely used in high energy hadronic interaction to obtain an information about
the geometric properties of the source. Multi-pion and photon spectra provide precise
information about reaction space time geometry in hadron-hadron and heavy ion collisions
\cite{Wiedemann:1999qn,Weiner:1999th}.

Another concept of momentum representation --- let call it ``momentum distribution'' ---
means to expand a wave function into stationary plane wave states \eqref{pln:wv:st}. It
was shown in Eq. \eqref{mom:mv:ev} that this was possible for even-$N$ states and
impossible for odd-$N$ states.

Such a momentum distribution in the ``infinite Universe'' would mean that
\begin{equation}\label{wv:fn:dstr}
    \Psi(\vec r,t)=\int d^3p\, \Phi(\vec p,t)\,e^{\frac{i}{\hbar}\vec p\cdot\vec
    r}\,e^{-i\frac{p^2}{2m\hbar}t}\,,
\end{equation}
what is in general not possible.

\begin{acknowledgments}
This paper is dedicated to Professor Jan {\L}opusza\'nski on his 80th birthday. Author is
indebted to Professor {\L}opusza\'nski for his whole life attitude which is a good
example of wisdom, courage, and a sense of humor.

I am grateful to R. Olkiewicz for interesting and inspiring discussions.

This work is partially supported by the Polish Committee for Scientific Research under
contract KBN~2~P03B~069~25.
\end{acknowledgments}
\break
\bibliography{sqw3-hep}

\begin{thebibliography}{31}
\expandafter\ifx\csname natexlab\endcsname\relax\def\natexlab#1{#1}\fi
\expandafter\ifx\csname bibnamefont\endcsname\relax
  \def\bibnamefont#1{#1}\fi
\expandafter\ifx\csname bibfnamefont\endcsname\relax
  \def\bibfnamefont#1{#1}\fi
\expandafter\ifx\csname citenamefont\endcsname\relax
  \def\citenamefont#1{#1}\fi
\expandafter\ifx\csname url\endcsname\relax
  \def\url#1{\texttt{#1}}\fi
\expandafter\ifx\csname urlprefix\endcsname\relax\def\urlprefix{URL }\fi
\providecommand{\bibinfo}[2]{#2}
\providecommand{\eprint}[2][]{\url{#2}}

\bibitem[{\citenamefont{Berry}(1996)}]{berr}
\bibinfo{author}{\bibfnamefont{M.~V.} \bibnamefont{Berry}},
  \bibinfo{journal}{J. Phys. A} \textbf{\bibinfo{volume}{29}},
  \bibinfo{pages}{6617} (\bibinfo{year}{1996}).

\bibitem[{\citenamefont{W\'ojcik et~al.}(2000)\citenamefont{W\'ojcik,
  Bia{\l}ynicki-Birula, and {\.Z}yczkowski}}]{wojc}
\bibinfo{author}{\bibfnamefont{D.}~\bibnamefont{W\'ojcik}},
  \bibinfo{author}{\bibfnamefont{I.}~\bibnamefont{Bia{\l}ynicki-Birula}},
  \bibnamefont{and}
  \bibinfo{author}{\bibfnamefont{K.}~\bibnamefont{{\.Z}yczkowski}},
  \bibinfo{journal}{Phys. Rev. Lett.} \textbf{\bibinfo{volume}{55}},
  \bibinfo{pages}{5022} (\bibinfo{year}{2000}), \eprint{quant-ph/0005060}.

\bibitem[{\citenamefont{Hu et~al.}(1999)\citenamefont{Hu, Li, Liu, and
  Gu}}]{Hu:1999bj}
\bibinfo{author}{\bibfnamefont{B.}~\bibnamefont{Hu}},
  \bibinfo{author}{\bibfnamefont{B.}~\bibnamefont{Li}},
  \bibinfo{author}{\bibfnamefont{J.}~\bibnamefont{Liu}}, \bibnamefont{and}
  \bibinfo{author}{\bibfnamefont{Y.}~\bibnamefont{Gu}}, \bibinfo{journal}{Phys.
  Rev. Lett.} \textbf{\bibinfo{volume}{82}}, \bibinfo{pages}{4224}
  (\bibinfo{year}{1999}), \eprint{chao-dyn/9903006}.

\bibitem[{\citenamefont{Aronstein and Stroud{,}~Jr.}(1997)}]{aron}
\bibinfo{author}{\bibfnamefont{D.~L.} \bibnamefont{Aronstein}}
  \bibnamefont{and} \bibinfo{author}{\bibfnamefont{C.~R.}
  \bibnamefont{Stroud{,}~Jr.}}, \bibinfo{journal}{Phys. Rev. A}
  \textbf{\bibinfo{volume}{55}}, \bibinfo{pages}{4526} (\bibinfo{year}{1997}).

\bibitem[{\citenamefont{Robinett}(2000)}]{robin1999}
\bibinfo{author}{\bibfnamefont{R.~W.} \bibnamefont{Robinett}},
  \bibinfo{journal}{Am. J. Phys.} \textbf{\bibinfo{volume}{68}},
  \bibinfo{pages}{410} (\bibinfo{year}{2000}).

\bibitem[{\citenamefont{Verdeyen}(1995)}]{verd}
\bibinfo{author}{\bibfnamefont{J.~T.} \bibnamefont{Verdeyen}},
  \emph{\bibinfo{title}{{Laser Electronics}}} (\bibinfo{publisher}{Prentice
  Hall}, \bibinfo{address}{Upper Saddle River, NJ}, \bibinfo{year}{1995}),
  \bibinfo{edition}{3rd} ed.

\bibitem[{\citenamefont{Arakawa and Yariv}(1986)}]{arakyari}
\bibinfo{author}{\bibfnamefont{Y.}~\bibnamefont{Arakawa}} \bibnamefont{and}
  \bibinfo{author}{\bibfnamefont{A.}~\bibnamefont{Yariv}},
  \bibinfo{journal}{IEEE Journal of Quantum Electronics}
  \textbf{\bibinfo{volume}{QE-22}}, \bibinfo{pages}{1887}
  (\bibinfo{year}{1986}).

\bibitem[{\citenamefont{Antoine et~al.}(2001)\citenamefont{Antoine, Gazeau,
  Monceau, Klauder, and Penson}}]{antoin2001gm}
\bibinfo{author}{\bibfnamefont{J.-P.} \bibnamefont{Antoine}},
  \bibinfo{author}{\bibfnamefont{J.-P.} \bibnamefont{Gazeau}},
  \bibinfo{author}{\bibfnamefont{P.}~\bibnamefont{Monceau}},
  \bibinfo{author}{\bibfnamefont{J.~R.} \bibnamefont{Klauder}},
  \bibnamefont{and} \bibinfo{author}{\bibfnamefont{K.~A.}
  \bibnamefont{Penson}}, \bibinfo{journal}{J. Math. Phys.}
  \textbf{\bibinfo{volume}{42}}, \bibinfo{pages}{2349} (\bibinfo{year}{2001}),
  \eprint{math-ph/0012044}.

\bibitem[{\citenamefont{Bonneau et~al.}(2001)\citenamefont{Bonneau, Faraut, and
  Valent}}]{bonneau:2001zq}
\bibinfo{author}{\bibfnamefont{G.}~\bibnamefont{Bonneau}},
  \bibinfo{author}{\bibfnamefont{J.}~\bibnamefont{Faraut}}, \bibnamefont{and}
  \bibinfo{author}{\bibfnamefont{G.}~\bibnamefont{Valent}},
  \bibinfo{journal}{Am. J. Phys.} \textbf{\bibinfo{volume}{69}},
  \bibinfo{pages}{322} (\bibinfo{year}{2001}), \eprint{quant-ph/0103153}.

\bibitem[{\citenamefont{Garbaczewski and Karwowski}(2001)}]{garb}
\bibinfo{author}{\bibfnamefont{P.}~\bibnamefont{Garbaczewski}}
  \bibnamefont{and}
  \bibinfo{author}{\bibfnamefont{W.}~\bibnamefont{Karwowski}},
  \emph{\bibinfo{title}{Impenetrable barriers and canonical quantization}}
  (\bibinfo{year}{2001}), \eprint{math-ph/0104010}.

\bibitem[{\citenamefont{In{\"o}nu and Wigner}(1952)}]{Inonu:1952ct}
\bibinfo{author}{\bibfnamefont{E.}~\bibnamefont{In{\"o}nu}} \bibnamefont{and}
  \bibinfo{author}{\bibfnamefont{E.~P.} \bibnamefont{Wigner}},
  \bibinfo{journal}{Nuovo Cimento} \textbf{\bibinfo{volume}{9}},
  \bibinfo{pages}{705} (\bibinfo{year}{1952}).

\bibitem[{\citenamefont{Bargmann}(1954)}]{bargm}
\bibinfo{author}{\bibfnamefont{V.}~\bibnamefont{Bargmann}},
  \bibinfo{journal}{Ann. Math.} \textbf{\bibinfo{volume}{59}},
  \bibinfo{pages}{1} (\bibinfo{year}{1954}).

\bibitem[{\citenamefont{Hamermesh}(1960)}]{hamerm}
\bibinfo{author}{\bibfnamefont{M.}~\bibnamefont{Hamermesh}},
  \bibinfo{journal}{Ann. Phys., N. Y.} \textbf{\bibinfo{volume}{9}},
  \bibinfo{pages}{518} (\bibinfo{year}{1960}).

\bibitem[{\citenamefont{Levy-Leblond}(1963)}]{levy:1963}
\bibinfo{author}{\bibfnamefont{J.-M.} \bibnamefont{Levy-Leblond}},
  \bibinfo{journal}{Journ. Math. Phys.} \textbf{\bibinfo{volume}{4}},
  \bibinfo{pages}{776} (\bibinfo{year}{1963}).

\bibitem[{\citenamefont{Levy-Leblond}(1974)}]{levy:1974}
\bibinfo{author}{\bibfnamefont{J.-M.} \bibnamefont{Levy-Leblond}},
  \bibinfo{journal}{Riv. Nuovo Cimento} \textbf{\bibinfo{volume}{4}},
  \bibinfo{pages}{99} (\bibinfo{year}{1974}).

\bibitem[{\citenamefont{Giulini}(1996)}]{giul:1996}
\bibinfo{author}{\bibfnamefont{D.}~\bibnamefont{Giulini}},
  \bibinfo{journal}{Ann. of Phys.} \textbf{\bibinfo{volume}{249}},
  \bibinfo{pages}{222} (\bibinfo{year}{1996}).

\bibitem[{\citenamefont{Landau and Lifshitz}(1977)}]{land}
\bibinfo{author}{\bibfnamefont{L.~D.} \bibnamefont{Landau}} \bibnamefont{and}
  \bibinfo{author}{\bibfnamefont{E.~M.} \bibnamefont{Lifshitz}},
  \emph{\bibinfo{title}{Quantum Mechanics: Nonrelativistic Theory}}
  (\bibinfo{publisher}{Pergamon}, \bibinfo{address}{Oxford},
  \bibinfo{year}{1977}), \bibinfo{edition}{3rd} ed.

\bibitem[{\citenamefont{Cohen-Tannoudji
  et~al.}(1977)\citenamefont{Cohen-Tannoudji, Diu, and Lalo{\"e}}}]{coh}
\bibinfo{author}{\bibfnamefont{C.}~\bibnamefont{Cohen-Tannoudji}},
  \bibinfo{author}{\bibfnamefont{B.}~\bibnamefont{Diu}}, \bibnamefont{and}
  \bibinfo{author}{\bibfnamefont{F.}~\bibnamefont{Lalo{\"e}}},
  \emph{\bibinfo{title}{Quantum Mechanics}}, vol.~\bibinfo{volume}{1}
  (\bibinfo{publisher}{John Wiley and Sons}, \bibinfo{address}{New York -
  London - Sydney - Toronto}, \bibinfo{year}{1977}).

\bibitem[{\citenamefont{von Neumann}(1955)}]{neumann:mfqm}
\bibinfo{author}{\bibfnamefont{J.}~\bibnamefont{von Neumann}},
  \emph{\bibinfo{title}{Mathematical Foundations of Quantum Mechanics}}
  (\bibinfo{publisher}{Princeton University Press}, \bibinfo{year}{1955}),
  \bibinfo{note}{{\it{Mathematische Grundlagen der Quantenmechanik}} (Julius
  Springer, Berlin, 1932)}.

\bibitem[{\citenamefont{von Neumann}(1929)}]{neumann:slf}
\bibinfo{author}{\bibfnamefont{J.}~\bibnamefont{von Neumann}},
  \bibinfo{journal}{Math. Ann.} \textbf{\bibinfo{volume}{102}},
  \bibinfo{pages}{49} (\bibinfo{year}{1929}).

\bibitem[{\citenamefont{Dunford and Schwartz}(1958)}]{dunf}
\bibinfo{author}{\bibfnamefont{N.}~\bibnamefont{Dunford}} \bibnamefont{and}
  \bibinfo{author}{\bibfnamefont{J.~T.} \bibnamefont{Schwartz}},
  \emph{\bibinfo{title}{Linear Operators}} (\bibinfo{publisher}{Interscience
  Publishers}, \bibinfo{address}{New York - London}, \bibinfo{year}{1958}).

\bibitem[{\citenamefont{Naimark}(1969{\natexlab{a}})}]{naim}
\bibinfo{author}{\bibfnamefont{M.~A.} \bibnamefont{Naimark}},
  \emph{\bibinfo{title}{Linear Differential Operators}}
  (\bibinfo{publisher}{Nauka}, \bibinfo{address}{Moscow},
  \bibinfo{year}{1969}{\natexlab{a}}), \bibinfo{edition}{2nd} ed.,
  \bibinfo{note}{{In Russian}}.

\bibitem[{\citenamefont{Naimark}(1969{\natexlab{b}})}]{naim1}
\bibinfo{author}{\bibfnamefont{M.~A.} \bibnamefont{Naimark}}, chap.
  \bibinfo{chapter}{V \S\,18.2}, in  \cite{naim}
  (\bibinfo{year}{1969}{\natexlab{b}}).

\bibitem[{\citenamefont{Dunford and Schwartz}(1963)}]{dunf1}
\bibinfo{author}{\bibfnamefont{N.}~\bibnamefont{Dunford}} \bibnamefont{and}
  \bibinfo{author}{\bibfnamefont{J.~T.} \bibnamefont{Schwartz}},
  \emph{\bibinfo{title}{Self Adjoint Operators in Hilbert Space}},
  chap.~\bibinfo{chapter}{12}, vol.~\bibinfo{volume}{2} of  \cite{dunf}
  (\bibinfo{year}{1963}).

\bibitem[{\citenamefont{Naimark}(1969{\natexlab{c}})}]{naim2}
\bibinfo{author}{\bibfnamefont{M.~A.} \bibnamefont{Naimark}}, chap.
  \bibinfo{chapter}{III \S\,9}, in  \cite{naim}
  (\bibinfo{year}{1969}{\natexlab{c}}).

\bibitem[{\citenamefont{Riesz and Sz.-Nagy}(1972)}]{riesz}
\bibinfo{author}{\bibfnamefont{F.}~\bibnamefont{Riesz}} \bibnamefont{and}
  \bibinfo{author}{\bibfnamefont{B.}~\bibnamefont{Sz.-Nagy}},
  \emph{\bibinfo{title}{Le\c cons d'Analyse Fonctionnele}}
  (\bibinfo{publisher}{Akad\'emiai Kiad\'o}, \bibinfo{address}{Budapest},
  \bibinfo{year}{1972}).

\bibitem[{\citenamefont{Lassner et~al.}(1987)\citenamefont{Lassner, Lassner,
  and Trapani}}]{Lassner:1987}
\bibinfo{author}{\bibfnamefont{G.}~\bibnamefont{Lassner}},
  \bibinfo{author}{\bibfnamefont{G.~A.} \bibnamefont{Lassner}},
  \bibnamefont{and} \bibinfo{author}{\bibfnamefont{C.}~\bibnamefont{Trapani}},
  \bibinfo{journal}{J. Math. Phys.} \textbf{\bibinfo{volume}{28}},
  \bibinfo{pages}{174} (\bibinfo{year}{1987}).

\bibitem[{\citenamefont{Kaempffer}(1965)}]{kaem}
\bibinfo{author}{\bibfnamefont{F.~A.} \bibnamefont{Kaempffer}},
  \emph{\bibinfo{title}{{Concepts in Quantum Mechanics}}}
  (\bibinfo{publisher}{Academic Press}, \bibinfo{address}{{New York and
  London}}, \bibinfo{year}{1965}), \bibinfo{note}{{A}ppendix 7}.

\bibitem[{\citenamefont{Hanbury~Brown and Twiss}(1956)}]{hbt56}
\bibinfo{author}{\bibfnamefont{R.}~\bibnamefont{Hanbury~Brown}}
  \bibnamefont{and} \bibinfo{author}{\bibfnamefont{R.~Q.} \bibnamefont{Twiss}},
  \bibinfo{journal}{Nature (London)} \textbf{\bibinfo{volume}{178}},
  \bibinfo{pages}{1046} (\bibinfo{year}{1956}).

\bibitem[{\citenamefont{Wiedemann and Heinz}(1999)}]{Wiedemann:1999qn}
\bibinfo{author}{\bibfnamefont{U.~A.} \bibnamefont{Wiedemann}}
  \bibnamefont{and} \bibinfo{author}{\bibfnamefont{U.~W.} \bibnamefont{Heinz}},
  \bibinfo{journal}{Phys. Rept.} \textbf{\bibinfo{volume}{319}},
  \bibinfo{pages}{145} (\bibinfo{year}{1999}), \eprint{nucl-th/9901094}.

\bibitem[{\citenamefont{Weiner}(2000)}]{Weiner:1999th}
\bibinfo{author}{\bibfnamefont{R.~M.} \bibnamefont{Weiner}},
  \bibinfo{journal}{Phys. Rept.} \textbf{\bibinfo{volume}{327}},
  \bibinfo{pages}{249} (\bibinfo{year}{2000}), \eprint{hep-ph/9904389}.

\end{thebibliography}

\end{document}